\documentclass[aps,prb,reprint,superscriptaddress,longbibliography]{revtex4-2}

\usepackage[T1]{fontenc}
\usepackage[utf8]{inputenc}
\usepackage{amsmath,amssymb,bm}
\usepackage{graphicx}
\usepackage{siunitx}
\usepackage[colorlinks=true,citecolor=blue,urlcolor=blue,linkcolor=blue]{hyperref}
\usepackage{comment}
\usepackage{xcolor}
\definecolor{mcgreen}{RGB}{0,170,100}

\newcommand{\CaCuW}{\ensuremath{\mathrm{Ca}_2\mathrm{CuWO}_6}}
\newcommand{\SrCuW}{\ensuremath{\mathrm{Sr}_2\mathrm{CuWO}_6}}
\newcommand{\BaCuW}{\ensuremath{\mathrm{Ba}_2\mathrm{CuWO}_6}}
\newcommand{\SrCuTe}{\ensuremath{\mathrm{Sr}_2\mathrm{CuTeO}_6}}
\newcommand{\BaCuTe}{\ensuremath{\mathrm{Ba}_2\mathrm{CuTeO}_6}}
\newcommand{\SrCuTeW}{\ensuremath{\mathrm{Sr}_2\mathrm{CuTe}_{1-x}\mathrm{W}_{x}\mathrm{O}_6}}
\newcommand{\BaMnTe}{\ensuremath{\mathrm{Ba}_2\mathrm{MnTeO}_6}}
\newcommand{\BaMnW}{\ensuremath{\mathrm{Ba}_2\mathrm{MnWO}_6}}
\newcommand{\CuIon}{\ensuremath{\mathrm{Cu}^{2+}}}

\newcommand{\WIon}{\ensuremath{\mathrm{W}^{6+}}}

\begin{document}

\title{Quasi two dimensional magnetic structure of the triclinic double perovskite Ca$_2$CuWO$_6$}

\author{Bangye Qin}
\email{bangye.qin.19@ucl.ac.uk}
\affiliation{Department of Physics and Astronomy, University College London,
Gower Street, London WC1E 6BT, United Kingdom}
\affiliation{Department of Physics, Durham University,
South Road, Durham DH1 3LE, United Kingdom}

\author{Dmitry D. Khalyavin}
\affiliation{ISIS Facility, STFC Rutherford Appleton Laboratory,
Didcot, Oxfordshire OX11 0QX, United Kingdom}

\author{Xuan Liang}
\affiliation{Research Center for Materials Nanoarchitectonics (MANA),
National Institute for Materials Science (NIMS),
Namiki 1-1, Tsukuba, Ibaraki 305-0044, Japan}
\affiliation{Graduate School of Chemical Sciences and Engineering,
Hokkaido University, North 10 West 8, Kita-ku,
Sapporo, Hokkaido 060-0810, Japan}

\author{Kazunari Yamaura}
\affiliation{Research Center for Materials Nanoarchitectonics (MANA),
National Institute for Materials Science (NIMS),
Namiki 1-1, Tsukuba, Ibaraki 305-0044, Japan}
\affiliation{Graduate School of Chemical Sciences and Engineering,
Hokkaido University, North 10 West 8, Kita-ku,
Sapporo, Hokkaido 060-0810, Japan}

\author{Alexei A. Belik}
\affiliation{Research Center for Materials Nanoarchitectonics (MANA),
National Institute for Materials Science (NIMS),
Namiki 1-1, Tsukuba, Ibaraki 305-0044, Japan}

\author{Roger D. Johnson}
\affiliation{Department of Physics, Durham University,
South Road, Durham DH1 3LE, United Kingdom}
\affiliation{Department of Physics and Astronomy, University College London,
Gower Street, London WC1E 6BT, United Kingdom}

\date{\today}

\begin{abstract}
We present the antiferromagnetic ground state of the triclinic double perovskite \CaCuW\ solved by neutron powder diffraction, and analyze it in direct comparison with tetragonal \SrCuW. Below $T_\mathrm{N} \simeq 32$~K, the magnetic Bragg reflections of \CaCuW\ are indexed by the commensurate propagation vector $\mathbf{k}=(\tfrac{1}{2},\tfrac{1}{2},0)$ in its native $P\bar{1}$ cell. Rietveld refinement yields a collinear structure with equal-magnitude, antiparallel moments on the crystallographically inequivalent Cu1 and Cu2 sites and an ordered moment of $0.66(3)\,\mu_{\mathrm B}$ per Cu at 1.5~K. A direct comparison with \SrCuW\ is obscured by different crystallographic settings and orientations of the cooperative Jahn--Teller elongation axes. Hence, we introduce a common crystallographic supercell through which effects of symmetry lowering from tetragonal to triclinic in the double perovskite are explored: Apparently different magnetic propagation vectors map onto the same supercell wave vector, revealing a common magnetic structure stabilized by tungsten-mediated, second-neighbor interactions. Mean field calculations further show that tetragonal symmetry preserves the degeneracy of four magnetic $\mathbf{k}$-domains in \SrCuW, whereas the triclinic splitting of symmetry-related exchange pathways in \CaCuW, most strongly within the Cu2 network, selects a single $\mathbf{k}$-domain. These results establish the magnetic ground state of \CaCuW\ and show how symmetry breaking selects a given ordered state without changing the underlying magnetic motif of the Sr analogue.
\end{abstract}

\maketitle

\section{Introduction}

Double perovskites with general chemical formula $A_2BB'\mathrm{O}_6$ provide a versatile platform for studying how crystal chemistry controls magnetic interactions. Large differences in charge and or ionic size between the $B$ and $B'$ cations commonly stabilize rock-salt order, and one can select for either all-magnetic $B$ and $B'$ cations, or magnetic $B\mathrm{O}_6$ octahedra spatially separated by nominally nonmagnetic $B'\mathrm{O}_6$ units~\cite{VasalaKarppinen2015,Iwanaga1999,Anderson1993,Paul2013}. The latter case forces magnetic coupling to proceed through extended super-superexchange pathways rather than through the shorter $B$--O--$B$ interactions found in $AB$O$_3$ perovskites~\cite{Vasala2012,Katukuri2020PRL}. As a result, the exchange hierarchy depends not only on bond geometry and the orbital state of the magnetic ion, but also on the electronic configuration of the intervening $B'$ cations and additional distortions permitted by the lattice~\cite{Katukuri2020PRL,Kanungo2016PRB,Mustonen2020ChemMater}. Ordered double perovskites therefore offer simultaneous chemical and structural control of magnetic dimensionality, frustration, and long-range order~\cite{Lufaso2004,Morrow2014,Mustonen2018PRB,Liang2025PRB,Paul2013}.

The cuprate members $A_2\mathrm{Cu}B'\mathrm{O}_6$ are especially sensitive to this interplay because \CuIon\ carries a $3d^9$, $S=1/2$ electronic configuration and is Jahn-Teller active~\cite{Lufaso2004,Lufaso2006}. Cooperative elongation of $\mathrm{CuO}_6$ octahedra parallel to a common axis selects half-filled $d_{x^2-y^2}$ orbitals in the plane perpendicular to the elongation axes ~\cite{Vasala2012}. Any exchange interaction involving a doubly occupied $d_{3z^2-r^2}$ orbital is typically taken to be approximately zero \cite{Vasala2012,Mustonen2024ChemMater}. Hence, the cooperative Jahn-Teller distortions effectively concentrate the strongest magnetic interactions within a quasi-two-dimensional Cu network~\cite{Iwanaga1999,Lufaso2006,Walker2016}. The contrasting behavior of the isostructural tellurate and tungstate compounds illustrates the additional role of the nonmagnetic linker. Let $J_1$ denote an in-plane nearest-neighbor Cu--Cu exchange interaction, while $J_2$ denotes a diagonal second-neighbor exchange of the corresponding 2D square-lattice model. In \SrCuTe\ and \BaCuTe, $J_1$ is dominant and favors N\'eel order. To the contrary, the low-lying empty $5d$ states of \WIon\ enhance the Cu--O--W--O--Cu contribution to $J_2$, producing a $J_2$-dominated square-lattice regime and so-called Type II antiferromagnetic order in \SrCuW\ and \BaCuW~\cite{Koga2016,Babkevich2016,VasalaJPCM2014,Walker2016,Mustonen2019ChemComm,Katukuri2020PRL,Todate2007}. The sensitivity of the exchange network to the $B'$ cation is further demonstrated by the disorder-driven magnetic states in \SrCuTeW\ and by the different magnetic structures of the nearly isostructural Mn analogues \BaMnTe\ and \BaMnW~\cite{Mustonen2018NatCommun,Mustonen2018PRB,Fogh2022PRB,Mustonen2024ChemMater,Mutch2020PRM,Mustonen2020ChemMater,Hu2021}.

Within the tungstate cuprates, the size of the $A$-site cation provides a second means of modifying the magnetic lattice. \SrCuW\ crystallizes in the tetragonal space group $I4/m$, with the cooperative Jahn-Teller elongations aligned along the tetragonal $c$ axis and the dominant Cu exchange network lying in the $ab$ plane~\cite{Gateshki2003,Lufaso2006,VasalaPRB2014}. Neutron diffraction and inelastic neutron scattering measurements established a quasi-two-dimensional collinear antiferromagnetic ground state below $T_N \simeq 24$~K with wavevector $\mathbf{k}=(\tfrac12,0,\tfrac12)$ and a dominant second-neighbor exchange interaction~\cite{VasalaJPCM2014,Walker2016}. Replacing Sr by the smaller Ca ion increases the octahedral tilting, and introduces additional distortions that reduce the coordination volume of the $A$-site cation, while substantially lowering the crystal symmetry ~\cite{DAY2012,Liang2025PRB}. Indeed, \CaCuW\ was recently synthesised under high-pressure high-temperature conditions and shown to crystallize in triclinic $P\bar{1}$ symmetry, with two crystallographically inequivalent Cu sites, two inequivalent W sites, and strongly distorted Jahn-Teller active $\mathrm{CuO}_6$ octahedra whose Jahn-Teller axes lie along $[-101]$ in the published triclinic setting~\cite{Liang2025PRB}. The two Cu sites form interpenetrating distorted square lattices in the $(10\bar{1})$ plane, and the corresponding minimal in-plane magnetic model contains no less than six symmetry-inequivalent exchanges. As shown later in Fig.~\ref{fig:magnetic-exchange}(b), we use $J_{1\alpha}$ and $J_{1\beta}$
to label two nearest-neighbor Cu1--Cu2 couplings that connect the
interpenetrating Cu networks, while four second-neighbor couplings
act within the individual networks: $J_{2\alpha}$ and
$J_{2\gamma}$ connect Cu1--Cu1 pairs along the two
inequivalent in-plane directions, whereas
$J_{2\beta}$ and $J_{2\delta}$ connect the
analogous Cu2--Cu2 pairs. Density-functional theory calculations gave
$J_{1\alpha}=J_{1\beta}=9$~K,
$J_{2\alpha}=51$~K,
$J_{2\gamma}=54$~K,
$J_{2\beta}=50$~K, and
$J_{2\delta}=80$~K,
revealing both the dominance of
second-neighbor exchange and the substantially stronger rectangular
anisotropy of the Cu2 network~\cite{Liang2025PRB}.

Bulk measurements indicate that the dominant magnetic energy scale of \CaCuW\ remains closely related to that of the better-known Sr and Ba tungstates. Its magnetic susceptibility exhibits a broad maximum near 60~K, characteristic of low-dimensional correlations, followed by long-range antiferromagnetic order at $T_N \simeq 32$~K; a field-induced transition is also observed below $T_N$~\cite{VasalaJPCM2014,Mustonen2019ChemComm,Liang2025PRB}. Different theoretical approaches consistently identify dominant antiferromagnetic second-neighbor interactions of order 60~K~\cite{Liang2025PRB}, and hence one might conclude that the replacement of Sr by Ca does not significantly alter the W-mediated exchange mechanism~\cite{Walker2016,Katukuri2020PRL,Liang2025PRB}. Instead, this $A$-site cation substitution primarily transforms a tetragonal Cu square lattice with one crystallographic Cu site into a triclinic system containing two inequivalent Cu networks and several symmetry-distinct exchange pathways. The ordered magnetic structure of \CaCuW, and the manner in which this structural complexity modifies the ground state relative to \SrCuW, is yet to be determined.

\begin{figure}
  \includegraphics[width=0.47\textwidth]
  {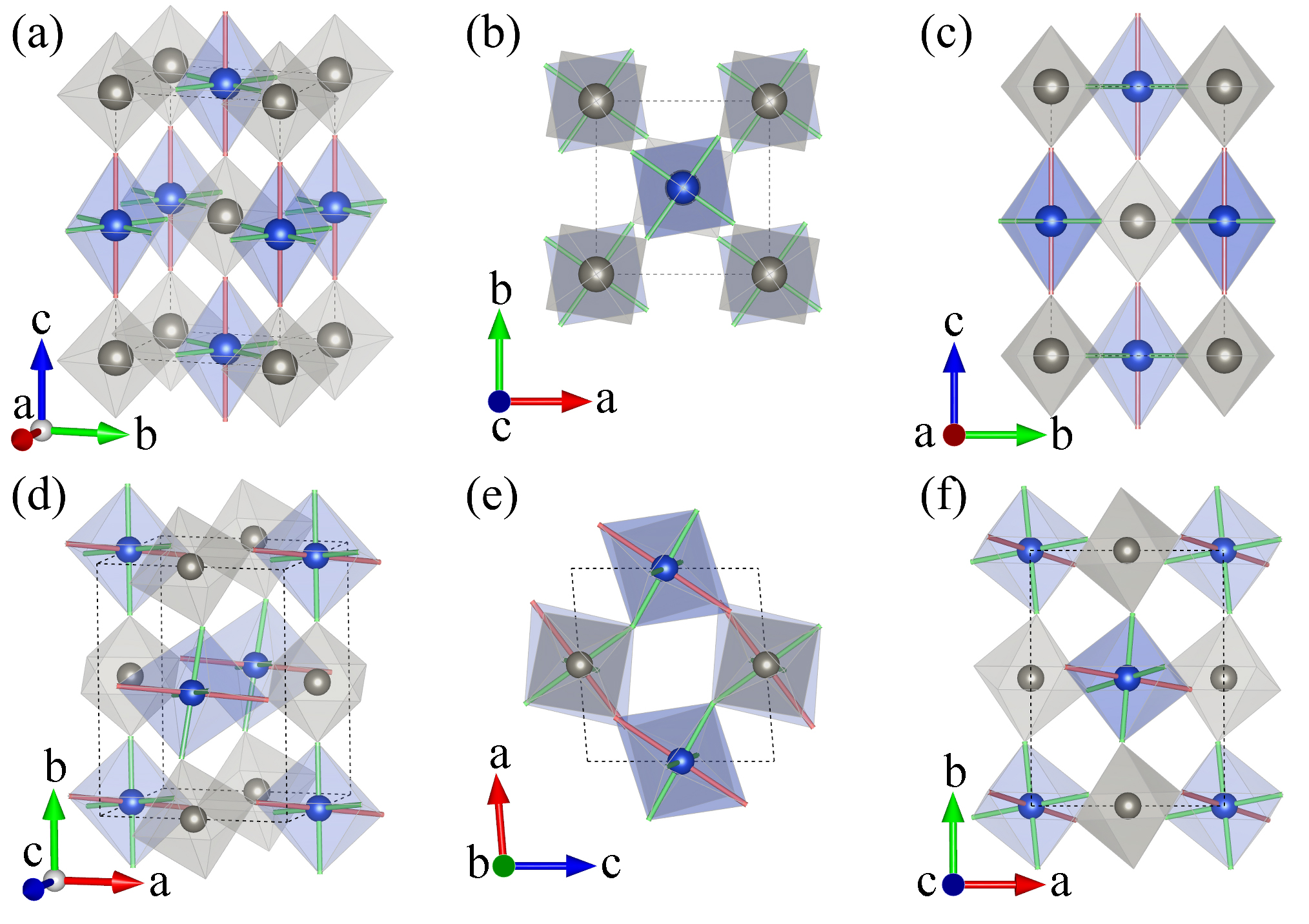}
  \caption{\label{fig:crystal-structures}
  Comparison of the crystal structures and co-operative Jahn--Teller
  distortions of \SrCuW\ and \CaCuW, shown in panes (a)-(c) and (d)-(f), respectively (note relative origin shift). The elongated Cu-O bonds shown in red are parallel to the tetragonal $c$ axis in \SrCuW\ and to
  $[\bar{1}01]$ in the native $P\bar{1}$ setting of \CaCuW, which follows Ref.~\cite{Liang2025PRB} in which $b$ is the largest unit cell parameter. W and Cu 
  atoms are shown in gray and blue, 
  respectively meanwhile Sr and Ca atoms are omitted for clarity.}
\end{figure}

In this paper we report a neutron powder diffraction study of the low-temperature crystal and magnetic structures of \CaCuW. Temperature-dependent refinements determine the antiferromagnetic ground state and the evolution of the ordered Cu moment through $T_N$. Mean-field calculations are then used to establish the role of symmetry-breaking within the magnetic exchange topology. Because the cooperative Jahn-Teller axes (and hence the quasi-2D-planes) align along different crystallographic directions in the conventional settings of the \CaCuW\ and \SrCuW\ crystal structures (see Figure~\ref{fig:crystal-structures}), a common supercell is introduced as a basis for comparing their exchange networks and real-space spin arrangements. This comparison separates differences arising from crystallographic convention from those produced by genuine symmetry lowering: We show that the two compounds retain the same W-mediated, second-neighbor-dominated magnetic motif, whereas triclinic \CaCuW\ contains inequivalent Cu networks and symmetry-split exchange pathways, absent in the tetragonal analogue, which select for a single $\mathbf{k}$-domain. Together, the neutron powder diffraction data analysis and theoretical model complete the magnetic characterization of \CaCuW\ and clarify how Jahn--Teller orbital order, octahedral tilting, and extended Cu--O--W--O--Cu exchange cooperate in this family of cuprate double perovskites.

\section{Methods}

\subsection{Sample synthesis}

A polycrystalline sample of \CaCuW\ with a total
mass of approximately 2~g was synthesized under high-pressure and
high-temperature conditions using a belt-type apparatus. A stoichiometric mixture of CuO, $\mathrm{WO_3}$, and pre-reacted $\mathrm{Ca_3WO_6}$ was thoroughly ground, loaded into Au capsules, and
treated at 6~GPa and approximately 1500~K for 2~h. At the end of the
heat treatment, the sample was quenched to room temperature, after which the pressure was slowly released. The $\mathrm{Ca_3WO_6}$ precursor was prepared from stoichiometric
amounts of $\mathrm{CaCO_3}$ and $\mathrm{WO_3}$ by annealing in air at
1270~K for 96~h with several intermediate grindings, following the
procedure reported in Ref.~\cite{Liang2025PRB}.

\subsection{Magnetometry}

DC magnetic susceptibility measurements were performed on
a polycrystalline \CaCuW\ sample using a superconducting
quantum interference device magnetometer (Quantum Design MPMS3). The
temperature-dependent susceptibility was measured between 2 and 400~K under a zero-field-cooled (ZFC) protocol under \(H = 10\,\mathrm{kOe}\). We note that the nominal sample mass was not corrected for the minor nonmagnetic impurity fraction and no diamagnetic corrections were applied to raw data.

\subsection{Neutron powder diffraction}

Neutron powder diffraction measurements were carried out using
the WISH cold-neutron time-of-flight diffractometer at ISIS Neutron
and Muon Source, UK
\cite{Chapon2011WISH,WISHInstrument}. Approximately 2~g of \CaCuW\ powder was loaded into a cylindrical vanadium can, and diffraction patterns were measured at 1.5~K and between 5 and 40~K in 5~K steps. The time-of-flight data were normalized, focused, and converted to diffraction histograms using the standard WISH data reduction workflow. Rietveld refinements of the nuclear and magnetic structures were
performed using the \textsc{FullProf} suite
\cite{RodriguezCarvajal1993}. Selected detector-bank histograms centred at $2\theta = 152.827^\circ$ and
$58.330^\circ$ were used to provide complementary
sensitivity to long-$d$ magnetic reflections while maintaining higher-$Q$ nuclear Bragg reflections.

\subsection{Mean-field calculations}

The mean-field spin Hamiltonian and resepctive magnetic ground state of \CaCuW\ and \SrCuW\ were analyzed following the approach of Luttinger and Tisza
\cite{LuttingerTisza1946,Luttinger1951}. Magnetic interactions were described by the Heisenberg Hamiltonian,
\begin{equation}
  \mathcal{H}
  = \frac{1}{2}
    \sum_{n\alpha,m\beta}
    J_{\alpha\beta}(\mathbf{R}_{m}-\mathbf{R}_{n})
    \mathbf{S}_{n\alpha}\!\cdot\!\mathbf{S}_{m\beta},
  \label{eq:heisenberg}
\end{equation}
where $n,m$ label Bravais-lattice cells, $\alpha,\beta$ label the Cu
sites within the chosen cell, and positive $J$ denotes antiferromagnetic exchange. The summation is taken over all nearest and next-nearest neighbour interactions. For \CaCuW, exchange interaction values were taken from the density-functional theory results of Ref.~\cite{Liang2025PRB}, while for \SrCuW, the corresponding values were based on the inelastic neutron scattering results of Ref.~\cite{Walker2016}. In both cases, the exchange topology was transformed to a common supercell as described in detail, below. For each system, the Fourier-transformed exchange matrix was constructed
as
\begin{equation}
  J_{\alpha\beta}(\mathbf{k})
  = \sum_{\boldsymbol{\delta}}
    J_{\alpha\beta}(\boldsymbol{\delta})
    \exp\!\left(2\pi i\,\mathbf{k}\!\cdot\!\boldsymbol{\delta}\right),
  \label{eq:jq}
\end{equation}
where $\boldsymbol{\delta}
=\mathbf{R}_m+\mathbf{r}_\beta-(\mathbf{R}_n+\mathbf{r}_\alpha)$
is the displacement vector connecting Cu site $\alpha$ in a given unit
cell to an exchange-coupled Cu site $\beta$, expressed
in the lattice coordinates of the chosen supercell. The lowest energy eigenvalue of $J_{\alpha\beta}(\mathbf{k})$ was then evaluated for all $\mathbf{k}$ within the first Brillouin zone. Degenerate modes that fold onto the same supercell wave vector on account of the artificially enlarged cell were identified by their rotational and translational symmetry equivalence when analysed in reference to the native Ca and Sr crystallographic cells. The mean-field calculations were then used to identify the exchange-selected propagation-vector and respective $\mathbf{k}$-domains. Because the method imposes only the global, or weak, spin-length constraint, the physical fixed-length spin configuration was determined independently from the neutron-diffraction refinement rather than inferred from the soft-mode eigenvector alone.

\section{Results and Discussion}
\label{sec:results}

\subsection{Crystal structure and magnetic ordering in \CaCuW}
\label{sec:ca-structure-magnetism}

\begin{figure}
  \includegraphics[width=0.48\textwidth]
  {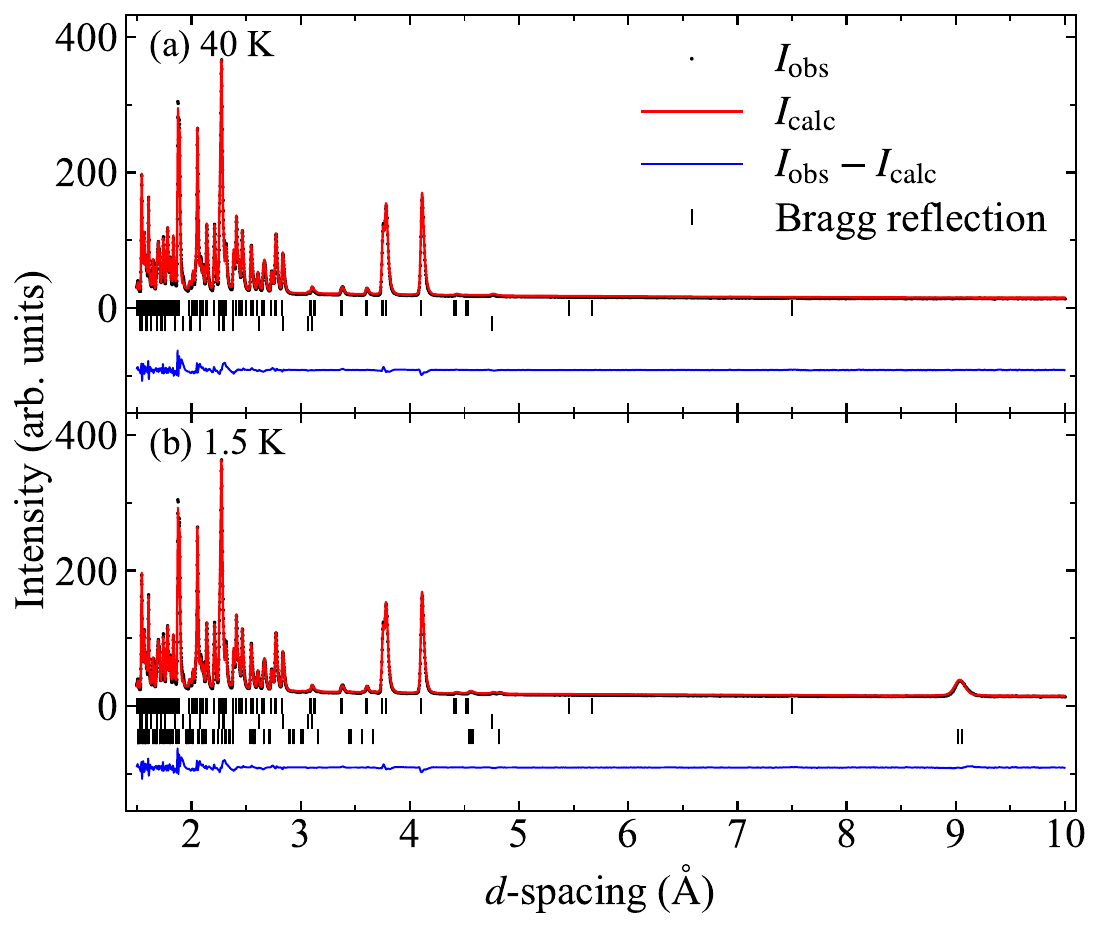}
  \caption{\label{fig:ca-npd}
  Neutron powder diffraction data for \CaCuW\ collected in (a) the paramagnetic phase and
  (b) the antiferromagnetic phase. A representative pattern from the detector bank centred at 58.330$^\circ$ is shown. Black points denote the observed intensities, the red curves are the calculated profiles, and the blue curves are the observed-minus-calculated differences, offset vertically for clarity. In (a), the tick marks indicate, from top to bottom, \CaCuW\ and $\mathrm{CaWO}_4$ impurity nuclear Bragg positions, respectively. In (b), the additional row indicates magnetic \CaCuW\ reflections. 
  The magnetic reflection near $d\simeq9.0$~\AA\ is indexed as $(\tfrac{1}{2},\tfrac{1}{2},0)$, uniquely identifying the propagation vector $\mathbf{k}=(\tfrac{1}{2},\tfrac{1}{2},0)$.}
\end{figure}

The previously reported triclinic crystal structure of \CaCuW\ \cite{Liang2025PRB} was used as a starting point in the following analysis, with the published setting of the lattice used throughout. In this setting, $b$ is the largest lattice parameter, and it is important to note that, unlike the tetragonal setting used for \SrCuW, this direction should not be confused with the local Jahn-Teller elongation axes that actually lie within the $ac$ plane. The \CaCuW\ nuclear structure was refined against neutron powder diffraction data measured within the paramagnetic phase at 40~K (see Figure \ref{fig:ca-npd}a). The reported $P\bar{1}$ model well accounts for the data, and the refined parameters are summarized in
Table~\ref{tab:ca-structure-40K}. The structure is in excellent agreement with that established by synchrotron X-ray powder diffraction at 295 K \cite{Liang2025PRB}, albeit with a cell volume approximately $0.49\%$ smaller; consistent with modest thermal contraction on cooling. Our refined model also contains an approximately $2.2$~wt.~\% $\mathrm{CaWO}_4$ impurity phase.

\begin{table*}
  \caption{\label{tab:ca-structure-40K}
  Crystallographic parameters of \CaCuW\ refined against neutron powder diffraction data measured at 40~K. The same space group setting as published in Ref.~\cite{Liang2025PRB} has been adopted. Note that atomic displacement parameters of a given element were constrained to be the same.}
  \squeezetable
  \begin{ruledtabular}
  \begin{tabular}{lcccccc}
    \multicolumn{6}{l}{\textbf{Unit-cell parameters}} \\
    $a$ (\AA) & 5.68439(9) & $\alpha$ (deg) & 90.117(1) \\
    $b$ (\AA) & 7.4980(1) & $\beta$ (deg) & 94.7552(8) \\
    $c$ (\AA) & 5.47925(9) & $\gamma$ (deg) & 90.234(1) \\
    $V$ (\AA$^3$) & 232.726(7)\\
    \\
    \multicolumn{6}{l}{\textbf{Atomic parameters}} \\
    Atom & Wyckoff site & $x$ & $y$ & $z$
      & $B_{\mathrm{iso}}$ (\AA$^2$) \\
    \colrule
    Ca1 & $2i$ & 0.0574(5) & 0.25396(73) & 1.0015(5)
        & 0.42(5) \\
    Ca2 & $2i$ & 0.5530(5) & 0.24774(74) & 0.5154(5)
        & 0.42(5) \\
    Cu1 & $1e$ & $\tfrac{1}{2}$ & $\tfrac{1}{2}$ & 0
        & 0.76(7) \\
    Cu2 & $1b$ & 0 & 0 & $\tfrac{1}{2}$
        & 0.76(7) \\
    W1  & $1d$ & $\tfrac{1}{2}$ & 0 & 0
        & 0.07(8) \\
    W2  & $1g$ & 0 & $\tfrac{1}{2}$ & $\tfrac{1}{2}$
        & 0.07(8) \\
    O1  & $2i$ & 0.4711(4) & 0.2469(6) & 0.0937(4)
        & 0.47(3) \\
    O2  & $2i$ & 0.9677(4) & 0.2506(6) & 0.4110(4)
        & 0.47(3) \\
    O3  & $2i$ & 0.3050(5) & 0.0452(5) & 0.7003(6)
        & 0.47(3) \\
    O4  & $2i$ & 0.7046(6) & 0.5476(4) & 0.3105(6)
        & 0.47(3) \\
    O5  & $2i$ & 0.2276(5) & 0.9443(4) & 0.1652(5)
        & 0.47(3) \\
    O6  & $2i$ & 0.8287(5) & 0.4555(4) & 0.7742(6)
        & 0.47(3) \\
    \\
    \multicolumn{6}{l}{\textbf{Refinement statistics}} \\
    \CaCuW\ (wt.~\%) & 97.76 & $R_p$ (\%) & 3.68 \\
    $\mathrm{CaWO}_4$ (wt.~\%) & 2.24 & $R_{\mathrm{exp}}$ (\%) & 0.72 \\
  \end{tabular}
  \end{ruledtabular}
\end{table*}

\begin{figure}
  \centering
  \includegraphics[width=0.47\textwidth]
  {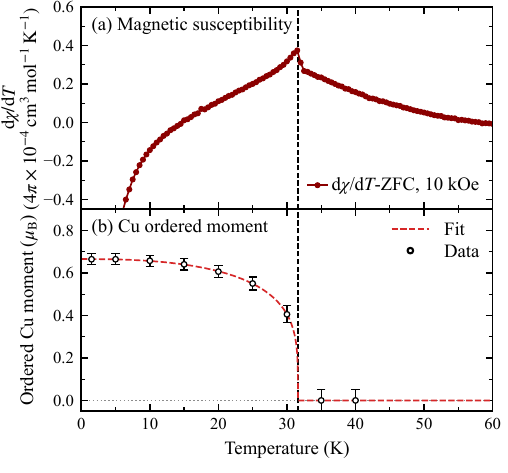}
  \caption{\label{fig:ca-moment}
  (a) Low-temperature dependence of
  the magnetic susceptibility for \CaCuW\ measured under zero-field-cooled (ZFC) conditions, shown here in SI units as the first derivative with respect to temperature to clearly identify the magnetic phase transition at $T_{\mathrm N}\simeq32$~K. (b) Temperature dependence of the ordered Cu moment per Cu site obtained from Rietveld refinements against NPD data. The dashed curve is the phenomenological fit described in the main text. The vertical dashed lines in both panels indicate the magnetic phase transition at $T_{\mathrm N}$.}
\end{figure}

Upon cooling to 1.5~K, additional Bragg intensities developed below $T_\mathrm{N} = 32$ K that are absent from the 40 K diffraction pattern (Fig.~\ref{fig:ca-npd}b), with the most prominent new reflection occurring at $d\simeq9.0$~\AA. This thermal evolution of the diffraction data is coincident with anomalous behaviour in the DC magnetic susceptibility shown in Figure \ref{fig:ca-moment}a, indicating that the low temperature intensities are magnetic in origin. A comprehensive propagation vector search throughout the first Brillouin zone of the $P\bar{1}$ crystal structure indexed the magnetic reflections with $\mathbf{k}=\left(\tfrac{1}{2},\tfrac{1}{2},0\right)$. This magnetic propagation vector doubles the periodicity along the $a$ and $b$ directions while leaving the translational periodicity along $c$ unchanged. 

\begin{figure}
  \centering
  \includegraphics[width=0.47\textwidth]
  {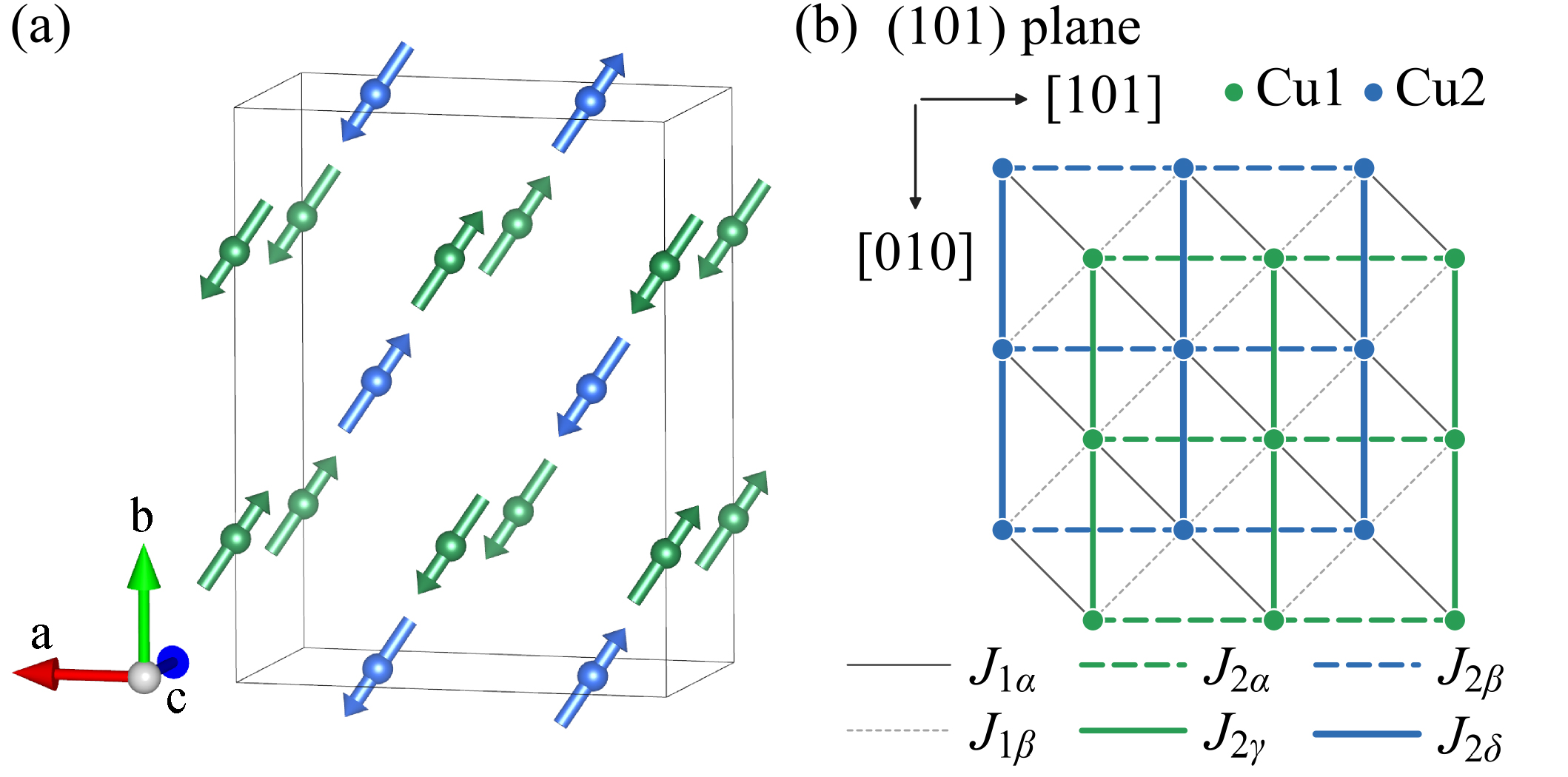}
  \caption{\label{fig:magnetic-exchange}
    (a) The collinear magnetic structure of \CaCuW\ at
    1.5~K, described by the propagation vector
    $\mathbf{k}=(\frac{1}{2},\frac{1}{2},0)$. The green spheres and arrows denote the Cu1 sites and their magnetic moments, respectively, while the blue spheres and arrows denote the Cu2 sites and their magnetic moments.
    (b) Projection of the quasi-2D Cu exchange network onto the distorted-square-lattice $(10\bar{1})$ plane. Cu1 and Cu2 form two interpenetrating rectangular sublattices. The dominant antiferromagnetic Cu1--Cu1 interactions ($J_{2\alpha}$ and $J_{2\gamma}$) and Cu2--Cu2 interactions ($J_{2\beta}$ and $J_{2\delta}$) act within the two sublattices, whereas the weaker $J_{1\alpha}$ and $J_{1\beta}$ interactions couple between them. The interplane pathways that select the $\mathbf{k}$-domain are omitted for clarity.}
\end{figure}

Two magnetic structure models were tested: First, Cu1 and Cu2 moments were constrained to be the same magnitude and sign, and with a global moment direction free to vary. Second, the Cu1 and Cu2 moments were constrained to be the same magnitude but opposite in sign, again with a freely varying global direction. Rietveld refinement against neutron powder diffraction data measured at 1.5 K (Figure \ref{fig:ca-npd}b) ruled out the former scenario, confirming antiparallel alignment of Cu1 and Cu2 moments in the ground state. As shown in Fig.~\ref{fig:magnetic-exchange}(a), the refined ordered moment at base temperature was $0.66(3)\,\mu_{\mathrm B}$ / Cu, aligned along the direction defined by the right-handed spherical angles $\phi = 122(8)^\circ,\theta =72(6)^\circ$, where $x\parallel a$, $y$ lying in the $ab$ plane, and $z\perp ab$ ($\phi$ is the azimuthal angle measured from $+x$ and $\theta$ is the polar angle measured from $+z$). The moment magnitude is significantly below the nominal $\sim 1~\mu_{\mathrm B}$ expected for an $S=1/2$ ion with $g\sim2$, indicating combined effects of quantum fluctuations in a quasi-two-dimensional antiferromagnet and Cu--O covalency {\cite{Walters2009,Sandvik1997,Sandvik2026,Mazurenko2015,Babkevich2016,Walker2016}}; the present powder-diffraction data do not separate these contributions.

The magnetic moment magnitude was refined against diffraction data measured as a function of temperature, and is plotted in Figure \ref{fig:ca-moment}b. The ordered moment was found to grow continuously on cooling below $T_{\mathrm N}$ and is essentially saturated below approximately 10~K. To parameterize this temperature dependence, we used the phenomenological function
\begin{equation}
  m_{\mathrm{Cu}}(T)=m_0
  \left[1-\left(\frac{T}{T_{\mathrm N}}\right)^{\alpha}\right]^{\beta},
  \qquad T<T_{\mathrm N},
  \label{eq:ca-order-parameter}
\end{equation}
with $m_{\mathrm{Cu}}=0$ for $T\geq T_{\mathrm N}$. A four-parameter fit gives $m_0=0.67(2)\,\mu_{\mathrm B}$/Cu and
$T_{\mathrm N}=32(5)$~K, consistent with the values given above. The shape parameters, $\alpha\sim2.5$ and $\beta\sim0.24$, are strongly covariant because of the sparse temperature sampling close to $T_{\mathrm N}$. Consequently, the dashed line in Fig.~\ref{fig:ca-moment}b is used only as a guide to the saturation of the ordered moment, and $\beta$ should not be interpreted as a critical exponent.

\subsection{$\mathbf{2\times2\times2}$ supercell}
\label{sec:common-supercell-lt}

As introduced above, in tetragonal \SrCuW\ the cooperative Jahn-Teller axes coincided with $\mathbf{c}_{\rm tet}$, while in triclinic \CaCuW\ the corresponding direction is
$[\bar{1}01]_{\rm tri}$. We therefore introduce a primitive
$2\times2\times2$ supercell (dimensions relative to the $AB$O$_3$ primitive unit cell), with Jahn-Teller axes parallel to $\mathbf{c}_{\rm sc}$. This supercell serves as a reference parent structure to both \SrCuW\ and \CaCuW. We define,

\begin{equation}
\begin{split}
\begin{pmatrix}
\bm{a}_{\rm sc}\\ \bm{b}_{\rm sc}\\ \bm{c}_{\rm sc}
\end{pmatrix}
&=
\underbrace{\begin{pmatrix}
1&1&0\\[-1mm]
-1&1&0\\[-1mm]
0&0&1
\end{pmatrix}}_{\mathsf S_{\rm tet}}
\begin{pmatrix}
\bm{a}_{\rm tet}\\ \bm{b}_{\rm tet}\\ \bm{c}_{\rm tet}
\end{pmatrix}, \quad \mathrm{and}\\[1mm]
\begin{pmatrix}
\bm{a}_{\rm sc}\\ \bm{b}_{\rm sc}\\ \bm{c}_{\rm sc}
\end{pmatrix}
&=
\underbrace{\begin{pmatrix}
1&0&1\\[-1mm]
0&1&0\\[-1mm]
-1&0&1
\end{pmatrix}}_{\mathsf S_{\rm tri}}
\begin{pmatrix}
\bm{a}_{\rm tri}\\ \bm{b}_{\rm tri}\\ \bm{c}_{\rm tri}
\end{pmatrix}.
\end{split}
\label{eq:common-cell-transform}
\end{equation}
Both transformations, $\mathsf{S}$, have determinant two, consistent with a doubling of the respective native unit cells. Further, we note that this construction is an exact change of crystallographic basis and does not impose cubic metric symmetry. Choosing appropriate unit cell origin shifts we define the Cu supercell fractional coordinates as
\begin{equation}
\begin{split}
\mathbf{r}_1&=[\tfrac12,0,0],\qquad
\mathbf{r}_2=[0,\tfrac12,0],\\
\mathbf{r}_3&=[0,0,\tfrac12],\qquad
\mathbf{r}_4=[\tfrac12,\tfrac12,\tfrac12].
\end{split}
\label{eq:common-cu-basis}
\end{equation}

In \SrCuW\, the four Cu sites form a single supercell site orbit, $\mathcal O^\mathrm{Sr}=\{\mathrm{Cu1},\mathrm{Cu2},\mathrm{Cu3},\mathrm{Cu4}\}$, while in \CaCuW\ the two crystallographically inequivalent Cu sites generate two supercell site orbits, $\mathcal O^\mathrm{Ca}_1=\{\mathrm{Cu1},\mathrm{Cu3}\}$ and $\mathcal O^\mathrm{Ca}_2=\{\mathrm{Cu2},\mathrm{Cu4}\}$.
Here \emph{orbit} is used in the crystallographic sense: each $\mathcal O_i$ is the set of supercell sites descended from one Cu sublattice. Wavevectors transform in the same way as the lattice basis vectors, \emph{i.e.} $\mathbf{k}_{\rm sc}= \mathsf{S}\mathbf{k}$. In tetragonal \SrCuW\ there exists 4 $\mathbf{k}$-domains related by the 4-fold rotation symmetry broken by the wavevector at the magnetic phase transition. They transform as
\begin{equation}
\begin{split}
\mathbf{k}_1 &= (0,\tfrac12,\tfrac12)_{\rm tet}
\longrightarrow(\tfrac12,\tfrac12,\tfrac12)_{\rm sc},\\
\mathbf{k}_2 &= (-\tfrac12,0,\tfrac12)_{\rm tet}
\longrightarrow(-\tfrac12,\tfrac12,\tfrac12)_{\rm sc},\\
\mathbf{k}_3 &= (0,-\tfrac12,\tfrac12)_{\rm tet}
\longrightarrow(-\tfrac12,-\tfrac12,\tfrac12)_{\rm sc},\\
\mathbf{k}_4 &= (\tfrac12,0,\tfrac12)_{\rm tet}
\longrightarrow(\tfrac12,-\tfrac12,\tfrac12)_{\rm sc}.
\end{split}
\label{eq:sr-k-transform}
\end{equation}
For triclinic \CaCuW, our neutron powder diffraction data identified a single wavevector that transforms as, 
\begin{equation}
\begin{split}
\mathbf{k}_3 &= (\tfrac12,\tfrac12,0)_{\rm tri}
\longrightarrow(\tfrac12,\tfrac12,-\tfrac12)_{\rm sc}
\end{split}
\label{eq:ca1-k-transform}
\end{equation}
where for consistency of labeling we have invoked the equivalence of $\pm\mathbf{k}$. One finds that the \CaCuW\ magnetic wavevector is identical to the \SrCuW\ wavevector, albeit one of four $\mathbf{k}$-domains of the tetragonal system. In reference to the four supercell wavevectors found in Equation \ref{eq:ca1-k-transform}, we identify three alternative wavevectors for \CaCuW\ through the inverse transformation.
\begin{equation}
\begin{split}
\mathbf{k}_1 &= (0,\tfrac12,\tfrac12)_{\rm tri},\\
\mathbf{k}_2 &= (-\tfrac12,\tfrac12,0)_{\rm tri},\\
\mathbf{k}_4 &= (0,-\tfrac12,\tfrac12)_{\rm tri}.
\end{split}
\label{eq:ca2-k-transform}
\end{equation}
Wavevectors $\mathbf{k}_1$($\mathbf{k}_3$) and $\mathbf{k}_4$($\mathbf{k}_2$) are symmetry-equivalent in the triclinic unit cell, and while the $\mathbf{k}_1$,$\mathbf{k}_4$ pair has been ruled out by experiment, all four will be considered for completeness in the mean-field analysis that follows.

The four supercell wavevectors are symmetry equivalent, modulo an allowed reciprocal-lattice vector. Distinct magnetic domains are hence defined by the relative phase of the 4 Cu moments within the primitive supercell. Table~\ref{tab:lt-domain-map} makes this correspondence explicit: For both \SrCuW\ and \CaCuW, sites in the same orbit are related by translational symmetry, $\mathbf{R}$, in the native crystal lattice, and hence their relative phase should be properly fixed by $2\pi \mathbf{k}\cdot\mathbf{R}$, where $\mathbf{R}$ is the lattice translation. For \SrCuW, the relative phases of all Cu sites are therefore fixed, but in \CaCuW\ the relative orientation of spins in different orbits remains a degree of freedom. Both parallel and antiparallel configurations are considered and assigned to the symmetry equivalent pairs of wavevectors $\mathbf{k}_1$($\mathbf{k}_3$) and $\mathbf{k}_4$($\mathbf{k}_2$) such that the same four domains found for \SrCuW\ are obtained for \CaCuW\, as summarised in Table  \ref{tab:lt-domain-map}.

\begin{table*}
  \caption{\label{tab:lt-domain-map}
  Correspondence between the four supercell magnetic domains and those of the native tetragonal and triclinic structures. The signs $\sigma_i$ denote the collinear moment directions on Cu sites within the unit cell (note that overall sign reversal
  generates time-reversal domains considered implicitly).}
  \squeezetable
  \setlength{\tabcolsep}{4pt}
  \begin{ruledtabular}
  \begin{tabular}{c|cc|cc|cc}
    Domain & $\mathbf{k}_{\rm sc}$
      & $(\sigma_1,\sigma_2,\sigma_3,\sigma_4)_{\rm sc}$
      & $\mathbf{k}_{\rm tet}$ & $(\sigma_1,\sigma_3)_{\rm tet}$
      & $\mathbf{k}_{\rm tri}$ & $(\sigma_1,\sigma_4)_{\rm tri}$ \\
    \colrule
    1 & $(\tfrac12,\tfrac12,\tfrac12)$
      & $(+,+,+,-)$
      & $(0,\tfrac12,\tfrac12)$ & $(+,+)$
      & $(0,\tfrac12,\tfrac12)$ & $(+,-)$ \\
    2 & $(-\tfrac12,\tfrac12,\tfrac12)$
      & $(-,+,+,+)$
      & $(-\tfrac12,0,\tfrac12)$ & $(+,-)$
      & $(-\tfrac12,\tfrac12,0)$ & $(+,-)$ \\
    3 & $(-\tfrac12,-\tfrac12,\tfrac12)$
      & $(+,+,-,+)$
      & $(0,-\tfrac12,\tfrac12)$ & $(+,-)$
      & $(-\tfrac12,-\tfrac12,0)$ & $(+,+)$ \\
    4 & $(\tfrac12,-\tfrac12,\tfrac12)$
      & $(+,-,+,+)$
      & $(\tfrac12,0,\tfrac12)$ & $(+,+)$
      & $(0,-\tfrac12,\tfrac12)$ & $(+,+)$ \\
  \end{tabular}
  \end{ruledtabular}
\end{table*}

The supercell construction also exposes symmetry breaking of the exchange topology.  As summarized in
Table~\ref{tab:lt-exchange-map}, both compounds are reported to host dominant, approximately $180^\circ$ short-short (s-s) next-nearest-neighbour (NNN) interactions, and significant $90^\circ$ s-s nearest-neighbour (NN) interactions. Here, `s-s' refers to the Cu--O bonds involved in the interaction, indicating exchange mediated via singly-occupied magnetic $d_{x^2-y^2}$ orbitals. Interactions involving long Cu--O bonds, and hence fully-occupied non-magnetic $d_{3z^2-r^2}$ orbitals, are denoted long-short (l-s) and long-long (l-l), and while included in our model, are generally considered negligible. Tetragonal symmetry reduces the exchange topology to one s-s NN exchange interaction, $J_1$, and one s-s NNN  interaction, $J_2$. Negligible NN and NNN interactions are labeled $J_3$ and $J_4$, respectively (see Table \ref{tab:lt-exchange-map}). The triclinic distortion then splits the exchange topology where, for example, the $J_2$ interaction becomes four strong exchanges $J_{21}$, $J_{22}$, $J_{23}$, and $J_{24}$ acting separately within the two Cu orbit networks.  

\begin{table*}
  \caption{\label{tab:lt-exchange-map}
  Symmetry relationship between exchange-interactions of \SrCuW\ \cite{Walker2016} and \CaCuW\ \cite{Liang2025PRB} as derived from our supercell approach.  Positive values denote antiferromagnetic interactions. The labels s and l denote paths involving the short and long Cu--O bonds, respectively. The interactions denoted
  $\simeq0$ are symmetry allowed but were not quantitatively resolved.}
  \squeezetable
  \begin{ruledtabular}
  \begin{tabular}{lllll}
    Interaction geometry & \SrCuW\ (meV) & \CaCuW\ (meV)
      & Cu--Cu connection & Displacement vector, $\bm\delta$ \\
    \colrule
    $90^\circ$ s--s NN & $J_1=2.45$ & $J_{1\alpha}=0.78$
      & Cu1--Cu2, Cu3--Cu4 & $[\tfrac12,\tfrac12,0]$ \\
    & & $J_{1\beta}=0.78$
      & Cu1--Cu2, Cu3--Cu4 & $[-\tfrac12,\tfrac12,0]$ \\
    \colrule
    $180^\circ$ s--s NNN & $J_2=8.83$ & $J_{2\alpha}=4.39$
      & Cu1--Cu1, Cu3--Cu3 & $[1,0,0]$ \\
    & & $J_{2\beta}=4.31$
      & Cu2--Cu2, Cu4--Cu4 & $[1,0,0]$ \\
    & & $J_{2\gamma}=4.65$
      & Cu1--Cu1, Cu3--Cu3 & $[0,1,0]$ \\
    & & $J_{2\delta}=6.89$
      & Cu2--Cu2, Cu4--Cu4 & $[0,1,0]$ \\
    \colrule
    $90^\circ$ l--s NN & $J_3\simeq0.00$ & $J_{3\alpha}\simeq0$
      & Cu1--Cu3 & $[\tfrac12,0,\tfrac12]$ \\
    & & $J_{3\beta}\simeq0$
      & Cu2--Cu4 & $[\tfrac12,0,\tfrac12]$ \\
    & & $J_{3\gamma}\simeq0$
      & Cu1--Cu3 & $[-\tfrac12,0,\tfrac12]$ \\
    & & $J_{3\delta}\simeq0$
      & Cu2--Cu4 & $[-\tfrac12,0,\tfrac12]$ \\
    & & $J_{3\epsilon}\simeq0$
      & Cu1--Cu4, Cu2--Cu3 & $[0,\tfrac12,\tfrac12]$ \\
    & & $J_{3\zeta}\simeq0$
      & Cu1--Cu4, Cu2--Cu3 & $[0,-\tfrac12,\tfrac12]$ \\
    \colrule
    $180^\circ$ l--l NNN & $J_4=0.01$ & $J_{4\alpha}\simeq0$
      & Cu1--Cu1, Cu3--Cu3 & $[0,0,1]$ \\
    & & $J_{4\beta}\simeq0$
      & Cu2--Cu2, Cu4--Cu4 & $[0,0,1]$ \\
  \end{tabular}
  \end{ruledtabular}
\end{table*}

\subsection{Mean field analysis}
\label{sec:lt}

The mean field Hamiltonian was constructed as a $2\times2$ matrix in the basis of the two Cu orbits, yielding
\begin{equation}
\mathcal J_\eta=
\begin{pmatrix}
a_1+\eta d_1 & b+\eta c\\
b+\eta c & a_2+\eta d_2
\end{pmatrix},
\label{eq:ca-lt-reduced}
\end{equation}
where, using the exchange-interaction notation of
Table~\ref{tab:lt-exchange-map},
\begin{align}
a_1&=-2(J_{2\alpha}+J_{2\gamma}+J_{4\alpha}),
&d_1&=2(J_{3\alpha}-J_{3\gamma}),\nonumber\\[-1mm]
a_2&=-2(J_{2\beta}+J_{2\delta}+J_{4\beta}),
&d_2&=2(J_{3\beta}-J_{3\delta}),\nonumber\\[-1mm]
b&=-2(J_{1\alpha}-J_{1\beta}),
&c&=2(J_{3\epsilon}-J_{3\zeta}).
\label{eq:ca-lt-coefficients}
\end{align}
We have introduced the parameter $\eta$ to encode the $\mathbf{k}$-domain; $\eta=+1$ corresponds to the observed $\mathbf{k}_3,\mathbf{k}_2$ pair, while $\eta=-1$ corresponds to the alternative $\mathbf{k}_1,\mathbf{k}_4$ pair. Having shown that the exchange energy has a local minimum for all $\mathbf{k}_i$, the Hamiltonian is solved separately for each wavevector pair, where the eigenvector with lowest energy eigenvalue describes the first ordered symmetry-adapted mode across the two Cu orbits. Comparing the lowest energy eigenvalues between $\eta=\pm1$ further determines the selected wavevector.

The parameters $a_1$ and $a_2$ give the $\mathbf{k}$-averaged diagonal energies of the two orbits, while $d_1$ and $d_2$ give wavevector dependent energy shifts.  Similarly, the $b$ parameter described wavevector-independent mixing, while $c$ captures the $\mathbf{k}$ dependence. We note that in the higher symmetry tetragonal phase of \SrCuW\ all parameters are exactly zero except for $a_1 = a_2$. One can show that the lowest energy eigenvalue for a given $\eta$ is
\begin{equation}
\begin{split}
\lambda_-(\eta)=\frac12\biggl\{&a_1+a_2+\eta(d_1+d_2)\\
&-\sqrt{[a_1-a_2+\eta(d_1-d_2)]^2
+4(b+\eta c)^2}\biggr\}.
\end{split}
\label{eq:ca-lt-eigenvalue}
\end{equation}

\begin{figure}
  \centering
  \includegraphics[width=0.48\textwidth]
  {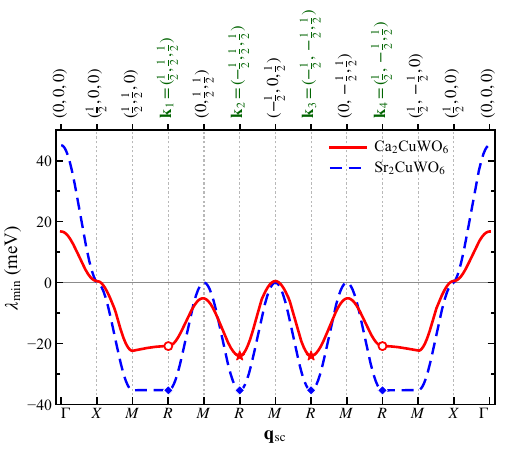} 
  \caption{\label{fig:energy}
	Lowest energy mean-field eigenvalue,
	$\lambda_{\min}$, for Ca$_2$CuWO$_6$ (red solid line) and Sr$_2$CuWO$_6$ (blue dashed line) plotted along a selected
	path in the first Brillouin zone of the pseudo-cubic supercell, as defined at the top of the figure. The four $R$ points correspond to the $\mathbf{k}$-domains (see Table~\ref{tab:lt-domain-map}) with $\mathbf{k}_{1,4}$ indicated by open circles and $\mathbf{k}_{2,3}$ by closed stars. We note that the nearly degenerate line of wavevectors appearing close to $\mathbf{k}_1$ and $\mathbf{k}_4$ in this plot result from the very weak, out-of-plane interactions $J_3$ and $J_4$ that involve doubly occupied $d_{3z^2-r^2}$ Cu orbitals.}
\end{figure}

We define
\begin{equation}
\Delta_k=\lambda_-(+1)-\lambda_-(-1),
\label{eq:ca-sector-splitting}
\end{equation}
for which $\Delta_k<0$ favors the experimentally observed $\mathbf{k}$-domain. One can show that for \SrCuW\ $\Delta_k=0$ and hence the four $\mathbf{k}$-domains remain degenerate by tetragonal symmetry, with the published exchange interaction values \cite{Walker2016} giving $\lambda_{\min}^{\rm Sr}=-35.34$~meV (see Fig. \ref{fig:energy}).

For \CaCuW, the reported exchange parameters
\cite{Liang2025PRB} split the two Cu mean-field modes for a
given wavevector yielding mean-field eigenvalues of $-18.10$ and
$-22.42$~meV, and hence uniquely determining a lower-energy mode which, within the mean-field weak spin-length constraint, is weighted towards the more strongly coupled Cu2 network. Surprisngly, one still finds that $\Delta_k=0$ because differences
among the weak $J_3$ interactions were not resolved.
Expanding Eqs.~\eqref{eq:ca-lt-eigenvalue} and
\eqref{eq:ca-sector-splitting} about these nondegenerate modes
to first order in $d_1$, $d_2$, and $c$ gives
\begin{equation}
\Delta_k \simeq d_1+d_2
-\frac{(a_1-a_2)(d_1-d_2)+4bc}
{\sqrt{(a_1-a_2)^2+4b^2}}.
\label{eq:ca-sector-selector}
\end{equation}
This expression gives an estimate of the $\mathbf{k}$-domain splitting of the lowest mean-field mode. For the reported exchange
parameters, $b=0$ and $a_2<a_1$, so
Eq.~\eqref{eq:ca-sector-selector} reduces to
$\Delta_k\simeq2d_2=4(J_{3\beta}-J_{3\delta})$.
Thus, $J_{3\beta}<J_{3\delta}$ would select the
experimentally observed $\mathbf{k}$-domain, as illustrated in Fig.~\ref{fig:energy}.

\section{Conclusions}
\label{sec:conclusions}

We have determined the antiferromagnetic ground state of
triclinic \CaCuW\ by neutron powder diffraction. Below
$T_{\mathrm N}\simeq32$~K, the magnetic Bragg reflections
are indexed by the commensurate propagation vector
$\mathbf{k}=(\frac12,\frac12,0)$ in the native
$P\overline{1}$ cell. Rietveld refinement gives a collinear
magnetic structure with equal-magnitude, antiparallel moments
on the crystallographically inequivalent Cu1 and Cu2 sites.
The ordered moment reaches $0.66(3)\,\mu_{\mathrm B}$ per Cu
at 1.5~K, significantly below the nominal spin-only value,
consistent with the combined effects of quantum fluctuations
in a quasi-two-dimensional antiferromagnet and Cu--O
covalency.

To compare this magnetic structure directly with that
of tetragonal \SrCuW, we introduced a common
crystallographic $2\times2\times2$ supercell.
Within this description, the apparently different
propagation vectors of the two compounds map onto the
same supercell wave vector modulo a reciprocal-lattice
vector. Furthermore, we showed that the magnetic structure of \CaCuW\
c6orresponds to one of the four magnetic
$\mathbf{k}$-domains of \SrCuW. The two compounds
therefore retain the same underlying magnetic motif,
stabilized primarily by strong, tungsten-mediated
second-neighbor antiferromagnetic interactions. As shown by our mean field analysis, the
principal effect of the triclinic distortion is
to split the exchange pathways, which allows weak interplane exchange to select for a single $\mathbf{k}$-domain. More generally, our results show how different magnetic energy scales can play distinct roles in establishing the magnetic ground state of double perovskites: Here, strong next-nearest-neighbour interactions define the underlying magnetic motif, while much weaker symmetry-breaking exchanges select between otherwise degenerate magnetic structures.

\section{Acknowledgments}
\label{sec:acknowledgments}

B.Q. acknowledges institutional support from University College London and a visiting studentship from Durham University. Neutron powder diffraction experiments were performed at the ISIS Neutron and Muon Source, Rutherford Appleton Laboratory, UK. This work was partially supported by a Grant-in-Aid for Scientific Research JP25K01657 from the Japan Society for the Promotion of Science.
  
\bibliography{references}

\end{document}